\documentclass{article}

\usepackage{arxiv}

\usepackage[utf8]{inputenc} 
\usepackage[T1]{fontenc}    
\usepackage{hyperref}       
\usepackage{url}            
\usepackage{booktabs}       
\usepackage{amsfonts}       
\usepackage{nicefrac}       
\usepackage{microtype}      
\usepackage{lipsum}		
\usepackage{graphicx}
\usepackage[style=ieee, citestyle=numeric-comp, backend=biber, dashed=false]{biblatex}
\usepackage{doi}
\usepackage{bm}
\usepackage{float}

\usepackage{amsmath,amssymb}

\DeclareMathOperator*{\EX}{\mathbb{E}}

\title{Inverse Design of Grating Couplers Using the Policy Gradient Method from Reinforcement Learning}

\author{ \href{https://orcid.org/0000-0003-1260-412X}{\includegraphics[scale=0.06]{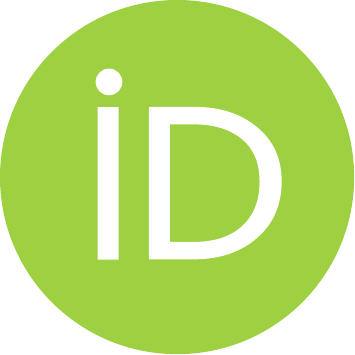}\hspace{1mm}Sean~Hooten}\thanks{Also affiliated with Department of Electrical Engineering and Computer Sciences, University of California, Berkeley, Berkeley, CA 94709, USA.}
		\\
	Hewlett Packard Labs \\
	Hewlett Packard Enterprise \\
	Milpitas, CA 95035, USA \\
	\texttt{shooten@eecs.berkeley.edu} \\
	\And
    Raymond G. Beausoleil \\
	Hewlett Packard Labs\\
	Hewlett Packard Enterprise \\
	Milpitas, CA 95035, USA \\
	\texttt{ray.beausoleil@hpe.com} \\
	\And
	\href{https://orcid.org/0000-0002-7301-8610}{\includegraphics[scale=0.06]{orcid.pdf}\hspace{1mm}Thomas~Van Vaerenbergh} \\
	Hewlett Packard Labs\\
	HPE Belgium\\
	B-1831 Diegem, Belgium \\
	\texttt{thomas.van-vaerenbergh@hpe.com}
}

\renewcommand{\shorttitle}{}

\hypersetup{
pdftitle={Inverse Design of Grating Couplers Using the Policy Gradient Method from Reinforcement Learning},
pdfsubject={physics, electromagnetics, optimization, reinforcement learning},
pdfauthor={Sean~Hooten, Thomas~Van Vaerenbergh},
pdfkeywords={artificial intelligence, reinforcement learning, machine learning, neural network, policy gradient, optimization, inverse design, adjoint method, physics, electromagnetics, photonics, grating coupler},
}

\begin{document}
\maketitle

\begin{abstract}
We present a proof-of-concept technique for the inverse design of electromagnetic devices motivated by the policy gradient method in reinforcement learning, named PHORCED (\underline{PH}otonic \underline{O}ptimization using \underline{R}EINFORCE \underline{C}riteria for \underline{E}nhanced \underline{D}esign). This technique uses a probabilistic generative neural network interfaced with an electromagnetic solver to assist in the design of photonic devices, such as grating couplers. We show that PHORCED obtains better performing grating coupler designs than local gradient-based inverse design via the adjoint method, while potentially providing faster convergence over competing state-of-the-art generative methods. As a further example of the benefits of this method, we implement transfer learning with PHORCED, demonstrating that a neural network trained to optimize 8$^\circ$ grating couplers can then be re-trained on grating couplers with alternate scattering angles while requiring \textgreater$10\times$ fewer simulations than control cases.
\end{abstract}


\section{Introduction}
There has been a recent, massive surge in research syncretizing topics in photonics and artificial intelligence / machine learning (AI/ML), including photonic analog accelerators \cite{shen_deep_2016, inagaki_coherent_2016, harris_quantum_2017, yamamoto_coherent_2017, khoram_nanophotonic_2018, hughes_wave_2019,  shastri_photonics_2021,  xu_11_2021}, physics emulators \cite{raissi_physics-informed_2019, gostimirovic_open-source_2019, guo_solving_2020, chen_physics-informed_2020, pestourie_active_2020, ghosh_machine_2021, lu_physics-informed_2021}, and AI/ML-enhanced inverse electromagnetic design techniques \cite{trivedi_data-driven_2019, qu_migrating_2019, melati_mapping_2019,tahersima_deep_2019, demeter-finzi_s-matrix_2019,dezfouli_design_2021,elsawy_global_2019, hammond_designing_2019,  jiang_global_2019, jiang_simulator-based_2020,jiang_multiobjective_2020, jiang_deep_2020,   hegde_deep_2020,   minkov_inverse_2020,  so_deep_2020,ma_parameter_2020,   kojima_deep_2021, lu_physics-informed_2021, ma_deep_2021,  melati_design_2021,  hegde_sample-efficient_2021}.  
While inverse electromagnetic design via local gradient-based optimization with the adjoint method has been successfully applied to a multitude of design problems throughout the entirety of the optics and photonics communities \cite{ jensen_topology_2011, lu_objective-first_2012, lalau-keraly_adjoint_2013, elesin_time_2014, piggott_inverse_2015,frellsen_topology_2016,  su_fully-automated_2017, michaels_leveraging_2018, michaels_inverse_2018, molesky_inverse_2018, hughes_adjoint_2018, liu_very_2018, andrade_inverse_2019, vercruysse_analytical_2019, augenstein_inverse_2020, bayati_inverse_2020,   hooten_adjoint_2020, jin_inverse_2018,jin_inverse_2020,  michaels_hierarchical_2020,  sun_adjoint-method-inspired_2020, lin_end--end_2021, omair_broadband_2021, vercruysse_inverse-designed_2021, zeng_inverse_2021}, inverse electromagnetic design leveraging AI/ML techniques promise superior computational performance, advanced data analysis and insight, or improved effort towards global optimization. For the lattermost topic in particular, Jiang and Fan recently introduced an unsupervised learning technique called GLOnet which uses a generative neural network interfaced with an electromagnetic solver to design photonic devices such as metasurfaces and distributed Bragg reflectors \cite{jiang_global_2019,jiang_simulator-based_2020,jiang_multiobjective_2020}. In this paper we propose a conceptually similar design technique, but with a contrasting theoretical implementation motivated by a concept in reinforcement learning called the policy gradient method -- specifically a one-step implementation of the REINFORCE algorithm \cite{williams_simple_1992, sutton_policy_2000}. We will refer to our technique as PHORCED = \underline{PH}otonic \underline{O}ptimization using \underline{R}EINFORCE \underline{C}riteria for \underline{E}nhanced \underline{D}esign. PHORCED is compatible with any external physics solver including EMopt \cite{michaels_emopt_2019}, a versatile electromagnetic optimization package that is employed in this work to perform 2D simulations of single-polarization grating couplers.

In Section\,\ref{sec:description}, we will qualitatively compare and contrast three optimization techniques: local gradient-based optimization (e.g., gradient ascent), GLOnet, and PHORCED. We are specifically interested in a proof-of-concept demonstration of the PHORCED optimization technique applied to grating couplers, which we present in Section\,\ref{sec:GCopt}. We find that both our implementation of the GLOnet method and PHORCED find better grating coupler designs than local gradient-based optimization, but PHORCED requires fewer electromagnetic simulation evaluations than GLOnet. Finally, in Section\,\ref{sec:TL} we introduce the concept of transfer learning to integrated photonic optimization, where in our application we demonstrate that a neural network trained to design 8$^\circ$ grating couplers with the PHORCED method can be re-trained to design grating couplers that scatter at alternate angles with greatly accelerated time-to-convergence. We speculate that a hierarchical optimization protocol leveraging this technique can be used to design computationally complex devices while minimizing computational overhead.

\section{Extending the Adjoint Method with Neural Networks}
\label{sec:description}

\begin{figure}[t!]
	\centering
	\includegraphics[width=6in]{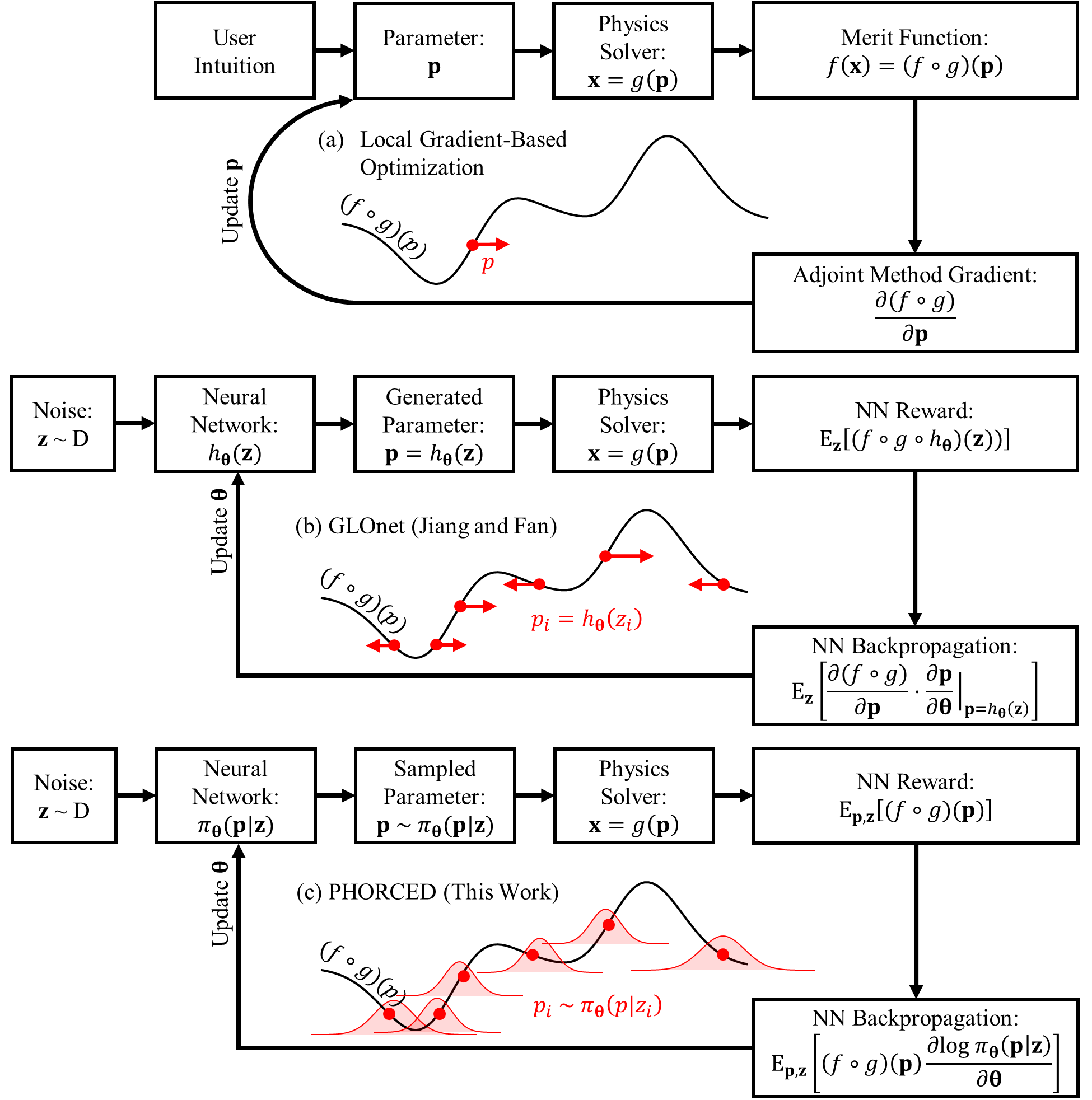}
	\caption{Neural networks provide a natural extension to conventional inverse design via the adjoint method. A typical gradient-based design loop is shown in (a) where the derivatives are calculated using the adjoint method. The GLOnet method (b), originally proposed by Jiang and Fan \cite{jiang_global_2019,jiang_simulator-based_2020}, replaces a conventional gradient-based optimization algorithm with a deterministic neural network. In this work we propose PHORCED (c), which uses a probabilistic neural network to generate devices. (b) and (c) are qualitatively similar, but require different gradients in backpropagation because of the representation of the neural network (deterministic versus probabilistic). In particular, notice that PHORCED does not require an evaluation of the adjoint method gradient of the electromagnetic merit function, $\frac{\partial (f\circ g)}{\partial \mathbf{p}}$.}
	\label{fig:fig1}
\end{figure}

Local gradient-based optimization using the adjoint method has been successfully applied to the design of a plethora of electromagnetic devices. Detailed tutorials of the adjoint method applied to electromagnetic optimization may be found in Refs.\,\cite{jensen_topology_2011, lu_objective-first_2012, lalau-keraly_adjoint_2013, elesin_time_2014,  michaels_leveraging_2018,   hughes_adjoint_2018, zeng_inverse_2021}. Here, we qualitatively illustrate a conventional design loop utilizing the adjoint method in Fig.\,\ref{fig:fig1}(a).  This begins with the choice of an initial electromagnetic structure parameterized by vector $\mathbf{p}$, which might represent geometrical degrees-of-freedom like the width and height of the device. This is fed into an electromagnetic solver, denoted by a function $g$. The resulting electric and magnetic fields, $\mathbf{x}=g(\mathbf{p})$, can be used to evaluate a user-defined electromagnetic \textit{merit function} $f(\mathbf{x})$ -- the metric that we are interested in optimizing (e.g., coupling efficiency). Gradient-based optimization seeks to improve the value of the merit function by updating the design parameters, $\mathbf{p}$, in a direction specified by the gradient of the electromagnetic merit function, $\frac{\partial (f\circ g)}{\partial \mathbf{p}}$. The value of the gradient may be obtained very efficiently using the adjoint method, requiring just two electromagnetic simulations regardless of the number of degrees-of-freedom (called the forward simulation and adjoint simulation, respectively).  A single iteration of gradient-based optimization is depicted visually in the center of Fig.\,\ref{fig:fig1}(a), where $p$ is a single-dimension point sampled along a toy merit function $(f\circ g)(p)$ representing the optimization landscape (which is unknown, a priori). The derivative (gradient) of the merit function is illustrated by the arrow pointing from $p$ in the direction of steepest ascent (assuming this is a maximization problem). During an optimization, we slowly update $p$ in this direction until a local optimum is reached.

The adjoint method chain-rule derivative of the electromagnetic merit function resembles the concept of \textit{backpropagation} in deep learning, where a neural network's weights can be updated efficiently by application of the chain-rule with information from the forward pass. Naturally, we might extend the functionality of the adjoint method by placing a neural network in the design loop. The neural network takes the place of a deterministic update algorithm (such as gradient ascent), potentially learning information or patterns in the design problem that allows it to find a better optimum. Below we present two methods to implement inverse design with neural networks: GLOnet (introduced by Jiang and Fan \cite{jiang_global_2019,jiang_simulator-based_2020}) and PHORCED (this work). Both methods are qualitatively similar, but differ in the representation of the neural network. In the main text of this manuscript we will qualitatively describe the differences between these techniques; a detailed mathematical discussion may be found in Supplementary Material Section 1.

The GLOnet optimization method is depicted qualitatively in Fig.\,\ref{fig:fig1}(b). The neural network is represented as a deterministic function $h_{\boldsymbol{\theta}}$ that takes in noise $\mathbf{z}$ from some known distribution $\text{D}$ and outputs design parameters $\mathbf{p}$. Importantly, the neural network is parameterized by programmable weights $\boldsymbol{\theta}$ that we intend to optimize in order to generate progressively better electromagnetic devices. Similar to regular gradient ascent, we may evaluate the electromagnetic merit function of a device generated by the neural network $(f\circ g)(\mathbf{p})$ using our physics solver, and find its gradient with respect to the design parameters using the adjoint method, $\frac{\partial (f\circ g)}{\partial \mathbf{p}}$. However, the GLOnet design problem is inherently stochastic because of the presence of noise, and therefore the optimization objective becomes the expected value of the electromagnetic merit function, $\EX_\mathbf{z}[(f\circ g\circ h_{\boldsymbol{\theta}})(\mathbf{z})]$ -- sometimes called the \textit{reward} in the reinforcement learning literature  \footnote{Note that we have written a generalized version of the reward function defined in the original works by Jiang and Fan \cite{jiang_global_2019, jiang_simulator-based_2020}. In that case, the reward function is chosen to weight good devices exponentially, i.e. $f\rightarrow \exp(f/\sigma)$ where $\sigma$ is a hyperparameter and $f$ is the electromagnetic quantity of interest. The full function is defined in Eq.\,(S.18) of Supplementary Materials Section 1.}. In practice this expression can be approximated by taking a simple average over the electromagnetic merit functions of several devices generated by the neural network per iteration. The gradient that is then backpropagated to the neural network is given by the expected value of the chain-rule gradient of the reward function. The first term, $\frac{\partial (f\circ g)}{\partial \mathbf{p}}$, can once again be computed very efficiently using the adjoint method, requiring just two electromagnetic simulations per device. Meanwhile the latter term, $\frac{\partial \mathbf{p}}{\partial \boldsymbol{\theta}}|_{\mathbf{p}=h_\mathbf{\theta}(\mathbf{z})}$, can be calculated internal to the neural network using conventional backpropagation with automatic differentiation. Visually, one iteration of the GLOnet method is shown in the center of Fig.\,\ref{fig:fig1}(b). In each iteration, the neural network in the GLOnet method suggests parameters, $p_i$, which are then individually simulated. Similar to gradient-based optimization from Fig.\,\ref{fig:fig1}(a), we find the gradient of the merit function value with respect to each generated design parameter, represented by the arrows pointing towards the direction of steepest ascent at each point. The net gradient information from many simulated devices effectively tells the neural network where to explore in the next iteration. With a dense search, the global optimum along the domain of interest can potentially be found.

Our technique, called PHORCED, is provided in Fig.\,\ref{fig:fig1}(c). Qualitatively speaking, it is very similar to GLOnet, but is motivated differently from a mathematical perspective. PHORCED is a special case of the REINFORCE algorithm \cite{williams_simple_1992, sutton_policy_2000} from the field of reinforcement learning (RL), applied to electromagnetic inverse design. In particular, the neural network is treated as purely probabilistic, defining a continuous conditional probability density function over parameter variables $\mathbf{p}$ conditioned on the input vector $\mathbf{z}$ -- denoted by $\pi_{\boldsymbol{\theta}}(\mathbf{p}|\mathbf{z})$. In other words, instead of outputting $\mathbf{p}$ deterministically given input noise, the neural network outputs probabilities of generating $\mathbf{p}$. We then randomly sample a parameter vector $\mathbf{p}$ for simulation and evaluation of the reward. Note that $\pi_{\boldsymbol{\theta}}(\mathbf{p}|\mathbf{z})$ is called the \textit{policy} in RL, and in this work is chosen to be a multivariate Gaussian distribution with mean vector and standard deviation of random variable $\mathbf{p}$ as outputs. The reward for PHORCED is qualitatively the same as GLOnet -- namely, we intend to optimize the expected value of the electromagnetic merit function. However, because both $\mathbf{p}$ and $\mathbf{z}$ are random variables, we take the joint expected value: $\EX_{\mathbf{p},\mathbf{z}}[(f\circ g)(\mathbf{p})]$.  Furthermore, because of the probabilistic representation of the neural network, the gradient of the reward with respect to the neural net weights for backpropagation is much different than the corresponding GLOnet case. In particular, we find that the backpropagated chain-rule gradient requires \textbf{no evaluation} of the gradient of the electromagnetic merit function, $\frac{\partial (f\circ g)}{\partial\mathbf{p}}$. Consequently, the electromagnetic adjoint simulation is no longer required, implying that fewer simulations are required overall for PHORCED compared to GLOnet under equivalent choices of neural network architecture and hyperparameters \footnote{However, because the representation of the neural network is different in either case, it would rarely make sense to use equivalent choices of neural network architecture and hyperparameters. Therefore, we make this claim tepidly, emphasizing only that we do not require adjoint simulations in the evaluation of the reward.}. PHORCED is visually illustrated in the center of Fig.\,\ref{fig:fig1}(c). The neural network defines Gaussian probability density functions conditioned on input noise vectors $z_i$, shown in the light red bell curves representing $\pi_\mathbf{\theta}(p|z_i)$, from which we sample points $p_i$ to simulate \footnote{Note that while we explored a uniform distribution at the input as well, our best results with PHORCED applied to grating coupler optimization in this work were attained with $\mathbf{z}$ drawn from a Dirac delta distribution, i.e. a constant vector input rather than noise. This has the effect of collapsing the multiple distributions depicted in Fig.\,\ref{fig:fig1}(c) into a single distribution from which we draw multiple samples.} Using information from the merit function values, the neural network learns to update the mean and standard deviation of the Gaussians. Consequently, we emphasize that the Gaussian policy distribution is not static because its statistical parameters are adjusted by the trainable weights of the neural network, and is therefore capable of exploring throughout the feasible design space. For adequate choice of distribution and dense enough search, the PHORCED method can potentially find the global optimum in the domain of interest. 

Before proceeding it should be remarked that the algorithms implemented by GLOnet and PHORCED have precedent in the literature, with some distinctions that we will outline here. Optimization algorithms similar to GLOnet were suggested in Refs.\,\cite{faury_neural_2018, faury_improving_2019}, where the main algorithmic difference appears in the definition of the reward function. In particular, the reward defined in Ref.\,\cite{faury_neural_2018} was the same generalized form that we have presented in Fig.\,\ref{fig:fig1}(b), while Jiang and Fan emphasized the use of an exponentially-weighted reward to enhance global optimization efforts \cite{jiang_simulator-based_2020}. On the other hand, PHORCED was motivated as a special case of the REINFORCE algorithm \cite{williams_simple_1992, sutton_policy_2000}, but also resembles some versions of evolutionary strategy \cite{hansen_cma_2016, salimans_evolution_2017, faury_neural_2018, faury_improving_2019}. The main difference between PHORCED and evolutionary strategy (ignoring several heuristics) is the explicit use of a neural network to model the multivariate Gaussian policy distribution, albeit some recent works have used neural networks in their implementations of evolutionary strategy \cite{salimans_evolution_2017,faury_improving_2019} for different applications than those studied here. Furthermore, PHORCED does not require a Gaussian policy; any explicitly-defined probability distribution can be used as an alternative if desired. Beyond evolutionary strategy, a recent work in fluid dynamics \cite{ghraieb_single-step_2021} uses an algorithm akin to PHORCED called One-Step Proximal Policy Optimization (PPO-1) -- a version of REINFORCE with a single policy, $\pi_{\boldsymbol{\theta}}$, operating on parallel instances of the optimization problem. The main distinction between PPO-1 and PHORCED is that we have implemented the option to use an input noise vector $\mathbf{z}$ to condition the output policy distribution, $\pi_{\boldsymbol{\theta}}(\mathbf{x}|\mathbf{z})$, which can effectively instantiate multiple distinct policies acting on parallel instances of the optimization problem. This potentially enables multi-modal exploration of the parameter space, bypassing a known issue of Bayesian optimization with Gaussian probability distributions \cite{faury_improving_2019}. However, note that the best results for the applications studied in this work used a constant input vector $\mathbf{z}$, thus making our implementation similar to PPO-1 in the results below. Regardless of the intricacies mentioned above, we emphasize that both GLOnet and PHORCED are unique in their application to electromagnetic optimization, to the authors' knowledge. In the next section we will compare all three algorithms from Fig.\,\ref{fig:fig1} applied to grating coupler optimization.

\section{Proof-of-Concept Grating Coupler Optimization}
\label{sec:GCopt}

A grating coupler is a passive photonic device that is capable of efficiently diffracting light from an on-chip waveguide to an external optical fiber. Recent works have leveraged inverse design techniques in the engineering of grating couplers, resulting in state-of-the-art characteristics \cite{michaels_inverse_2018, su_fully-automated_2017, hooten_adjoint_2020, sun_adjoint-method-inspired_2020, dezfouli_design_2021}. In this section we will show how generative neural networks can aid in the design of ultra-efficient grating couplers.

\begin{figure}[t!]
    \centering
    \includegraphics[width=6in]{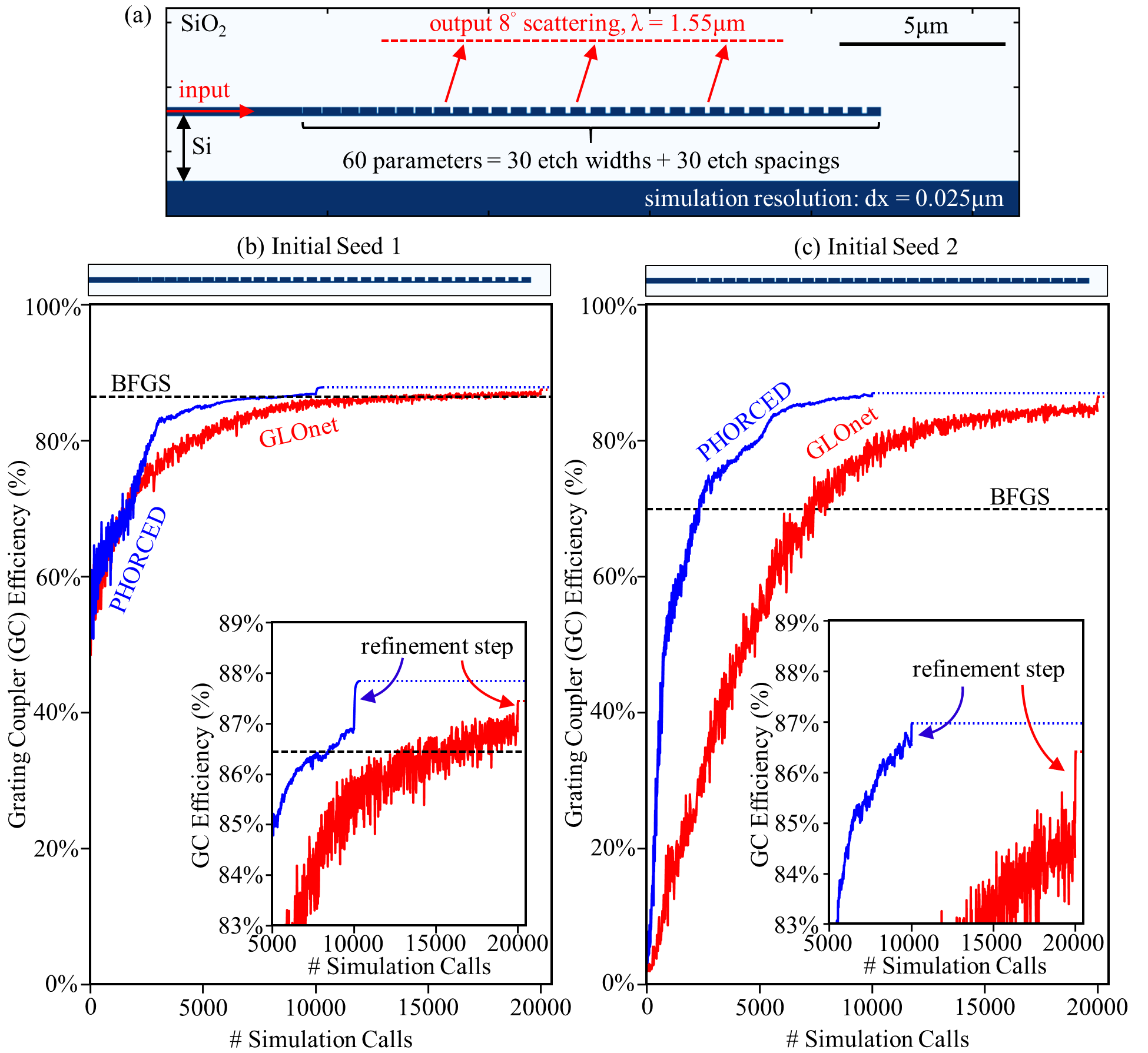}
    \caption{PHORCED and GLOnet outperform conventional gradient-based optimization, with different simulation evaluation requirements. The grating coupler simulation geometry for optimization is shown in (a), consisting of a SOI wafer platform with 280nm waveguide, 190nm etch depth, and 2\textmu m BOX height. We optimized 60 device parameters in total, namely the width and spacing between 30 etch corrugations. The results of the BFGS, GLOnet, and PHORCED algorithms applied to Initial Seed 1 and Initial Seed 2 are presented in (b) and (c) as a function of the number of simulation calls. The initial seeds (grating designs) are illustrated above each optimization plot. The insets depict zoomed-in views of the peak efficiencies attained by PHORCED and GLOnet, along with respective BFGS refinement steps performed on the best design generated by each algorithm.}
    \label{fig:fig2}
\end{figure}

The grating coupler geometry used for our proof-of-concept is depicted in Fig.\,\ref{fig:fig2}(a). We assume a silicon-on-insulator (SOI) wafer platform with a $280\,\mathrm{nm}$-thick silicon waveguide separated from the silicon substrate by a 2\textmu m buried oxide (BOX) layer. The grating coupler consists of periodically spaced corrugations to the input silicon waveguide with etch depth $190\,\mathrm{nm}$. These grating coupler dimensions are characteristic of a high-efficiency integrated photonics platform operating in the C-band ($1550\,\mathrm{nm}$ central wavelength) as suggested by a transfer matrix based directionality calculation \cite{michaels_hierarchical_2020,michaels_gcslab_2019}, and are provided as an example in the open-source electromagnetic optimization package EMopt \cite{michaels_emopt_2019} that was used to perform forward and adjoint simulations in this work. 

For our optimizations, we will consider 60 total designable parameters that define the grating coupler: the width of and spacing between 30 waveguide corrugations. For a well-designed grating coupler, input light to the waveguide scatters at some angle relative to vertical toward an external optical fiber. In this work we choose to optimize the grating coupler at fixed wavelength and scattering angle assuming specific manufacturing and assembly requirements for a modular optical transceiver application \cite{mathai_detachable_2020, hooten_adjoint_2020}. In our case, the merit function for optimization is the coupling efficiency of scattered light with wavelength $\lambda=1550\,\text{nm}$ propagating $8^\circ$ relative to normal, mode-matched to an output Gaussian beam mode field diameter 10.4\textmu m -- characteristic of a typical optical fiber mode. The explicit definition of this electromagnetic merit function may be found in Refs.\,\cite{michaels_inverse_2018,hooten_adjoint_2020,watanabe_perpendicular_2017}. Note that we did not include fabrication constraints nor other specifications of interest in grating couplers in our parameterization choice, e.g. the BOX thickness and the Si etch depth, which will be desirable in future optimizations of experimentally-viable devices. Furthermore, the simulation domain is discretized with a $dx=25\text{nm}$ grid step which may result in some inaccuracy for very fine grating coupler features. This simulation discretization was chosen for feasibility of the optimization since individual simulations require about 4 seconds to compute on a high-performance server with over 30 concurrent MPI processes, and as we will show GLOnet and PHORCED can require as many as 20,000 simulation evaluations for convergence. Nevertheless, we utilized permittivity smoothing \cite{michaels_leveraging_2018} to minimize the severity of this effect and obtain physically meaningful results.

We apply the Broyden–Fletcher–Goldfarb–Shanno (BFGS), GLOnet, and PHORCED algorithms to grating coupler optimization with two different initial designs (Initial Seed 1 and Initial Seed 2) in Fig.\,\ref{fig:fig2}(b) and Fig\,\ref{fig:fig2}(c). Initial Seed 1 corresponds to the grating depicted in Fig.\,\ref{fig:fig2}(a) and the top of Fig.\,\ref{fig:fig2}(b), where we used a parameter sweep to choose a linear apodization of the etch duty cycle before optimization. Initial Seed 2 corresponds to the grating shown at the top of Fig.\,\ref{fig:fig2}(c) where we use a uniform duty cycle of 90\%. Both initial designs have pitch that satisfy the grating equation \cite{michaels_inverse_2018, hooten_adjoint_2020} for $8^\circ$ scattering. These initial seeds serve to explore the robustness of the optimization algorithms to  ``good'' and ``poor'' choices of initial condition. Indeed, Initial Seed 1 satisfies physical intuition for a good grating coupler, because chirping the duty cycle is well-known to improve Gaussian beam mode-matching, and thus the initial grating coupler efficiency is already a reasonable value of 56\%. Meanwhile, Initial Seed 2 has the correct pitch for $8^\circ$ scattering, but has a low efficiency of 1\% owing to poor mode-matching and directionality. In essence, we are using these two cases as a proxy to explore whether PHORCED and GLOnet can reliably boost electromagnetic performance in a high-dimensional parameter space, even in cases where the intuition about the optimal initial conditions is limited.

BFGS is a conventional gradient-based optimization algorithm similar to that depicted in Fig.\,\ref{fig:fig1}(a), and implemented using default settings from the open-source SciPy optimize module. After optimization with BFGS, the final simulated grating coupler efficiency of Initial Seed 1 and Initial Seed 2 are 86.4\% and 69.9\% respectively, which are shown in the black dashed lines of Fig.\,\ref{fig:fig2}(b) and Fig.\,\ref{fig:fig2}(c). Note that the number of simulation calls for BFGS is not shown because it is vastly smaller than that required for GLOnet and PHORCED (144 and 214 total simulations for Initial Seed 1 and Initial Seed 2, respectively). 

The implementation of GLOnet and PHORCED for the grating coupler optimizations in Fig.\,\ref{fig:fig2}(b) and Fig.\,\ref{fig:fig2}(c) are described below; other details and specifications, such as a graphical illustration of the neural network models used in either case, may be found in Supplementary Materials Section 2. GLOnet is described qualitatively in Fig.\,\ref{fig:fig1}(b) where we use a deterministic neural network and an exponentially-weighted electromagnetic merit function originally recommended by Jiang and Fan \cite{jiang_global_2019,jiang_simulator-based_2020} with a chosen hyperparameter $\sigma=0.6$. PHORCED is described qualitatively in Fig.\,\ref{fig:fig1}(c) where we use a probabilistic neural network modeling a multivariate, isotropic Gaussian output distribution. The electromagnetic merit function used in the reward is just the unweighted grating coupler efficiency, except we used a ``baseline'' subtraction of the average merit function value in the backpropagated gradient (which is a common tactic in reinforcement learning for reducing model variance \cite{sutton2018reinforcement}). In both methods, we use a stopping criterion of 1,000 total optimizer iterations, with 10 devices sampled per iteration. Note that because GLOnet requires an adjoint simulation for each device and PHORCED does not, the effective stopping criteria are $1000\times10\times2=20,000$ simulation calls for GLOnet and $1000\times10\times1=10,000$ simulation calls for PHORCED. The neural network models for GLOnet and PHORCED were implemented in PyTorch, and are illustrated in Fig. S1 of the Supplementary Material. For GLOnet, we used a convolutional neural network with ReLU activations and linear output of the design vector. For PHORCED, we used a simple fully-connected neural network with ReLU activations and linear output defining the statistical parameters (mean and variance) of the Gaussian policy distribution. We applied initial conditions by adding the design vector representing Initial Seed 1 and Initial Seed 2 to either the direct neural network output or the mean of the Gaussian distribution for GLOnet and PHORCED, respectively. Hyperparameters such as the number of weights and learning rate of the neural network optimizer were individually tuned for GLOnet and PHORCED (specifications can be found in Fig. S2 of the Supplemental Material). For both neural networks, we used an input vector $\mathbf{z}$ of dimension 5. However, for GLOnet $\mathbf{z}$ was drawn from a uniform distribution $\mathcal{U}(-1,1)$, while for PHORCED we used a simple Dirac delta distribution centered on 1. In other words, the input for PHORCED was a constant vector of 1's. In this work we achieved our best results with this choice, but similar results (within 1\% absolute grating coupler efficiency) could be attained with alternative choices of input distribution. Anecdotally, we found that using a noisy input could improve training stability and performance in toy problems studied outside of this work, but further investigation is required.  

We find that PHORCED and GLOnet outperform regular BFGS for both initial conditions studied in Fig\,\ref{fig:fig2}. For Initial Seed 1 (Fig.\,\ref{fig:fig2}(b)), we generated optimized grating coupler efficiencies of 86.9\% and 87.2\% for PHORCED and GLOnet, respectively. For Initial Seed 2 (Fig.\,\ref{fig:fig2}(c)) we find optimized grating coupler efficiencies of 86.8\% and 85.6\% for PHORCED and GLOnet, respectively. Furthermore, as shown in the insets of Fig.\,\ref{fig:fig2}(b) and Fig.\,\ref{fig:fig2}(c), we were able to marginally improve each of the results by applying a BFGS ``refinement step'' to the best performing design output from GLOnet and PHORCED. This refinement step was limited to a maximum of 200 iterations, or until another default convergence criterion was met. For Initial Seed 1 we obtained improvements of $\{86.9\%\rightarrow87.8\%\}$/$\{87.2\%\rightarrow 87.4\%$) for PHORCED/GLOnet, respectively. For Initial Seed 2, we find improvements of {$\{86.8\%\rightarrow87.0\%\}$/ $\{85.6\%\rightarrow 86.4\%\}$} for PHORCED/GLOnet, respectively. Since PHORCED and GLOnet are inherently statistical and noisy, the refinement step is useful for finding the nearest optimum without requiring one to run an exhaustive search of the neural network generator in inference mode,.

In summary, we find that the PHORCED + BFGS refinement optimization achieved the best performance for both Initial Seed 1 and Initial Seed 2 with final grating coupler efficiency of 87.8\% and 87.0\%, respectively. These results agree with a transfer matrix based directionality analysis of these grating coupler dimensions \cite{michaels_hierarchical_2020, michaels_gcslab_2019}, where we find that approximately 88\% grating efficiency is possible under perfect mode-matching conditions -- meaning that our result for Initial Seed 1 is close to a theoretical global optimum. Notably, GLOnet had better performance than PHORCED in Initial Seed 1 before the refinement step was applied, and it is possible that better results could have been achieved with further iterations of both algorithms (as indicated by the slowly rising slopes of the optimization curves in the insets of Fig.\,\ref{fig:fig2}(b)-(c)). However, we emphasize that PHORCED required approximately $2\times$ fewer simulation than GLOnet with the same number of optimization iterations because of the lack of adjoint gradient calculations. 

Perhaps the most important result of these optimizations is that both PHORCED and GLOnet proved to be resilient against our choice of initial condition. Indeed, while BFGS provided a competitive result for Initial Seed 1, it failed to find a favorable local optimum given Initial Seed 2. PHORCED and GLOnet, on the other hand, attained final results within 1\% absolute grating coupler efficiency for both initial conditions. This outcome is promising when considering the relative sparsity of each algorithm's search in a high-dimensional design space. Indeed, as mentioned previously, we only sampled 10 devices per iteration of the optimization, meaning that there were fewer samples than dimension of the resulting parameter vectors (60). As an additional reference, we compared our results to CMA-ES (implemented with the open-source package pycma), a popular ``blackbox'' global optimization algorithm that is known to be effective for high-dimensional optimization \cite{hansen_cma_2016}. Under the same number of simulation calls as PHORCED (10,000), CMA-ES reached efficiencies of 87.3\% and 86.7\% for Initial Seed 1 and Initial Seed 2, respectively. Therefore our implementations of PHORCED and GLOnet are competitive with current state-of-the-art blackbox algorithms. While the simulation requirements for convergence of PHORCED and GLOnet remain computationally prohibitive for more complex electromagnetic structures than those studied here, in contrast to local gradient-based search, our results offer the possibility of global optimization effort in electromagnetic design problems, where we are capable of limiting the tradeoff of performance with search density as well as the need for human intervention in situations where physical intuition is more difficult to ascertain. Furthermore, we believe that our implementations of GLOnet and PHORCED have significant room for improvement, and have advantages that go beyond alternative global optimization algorithms like CMA-ES. In particular, leveraging advanced concepts in deep learning and reinforcement learning can further improve computational efficiency and performance. For example, whereas our implementations of PHORCED and GLOnet used simple neural networks, a ResNet architecture can improve neural network generalizability while simultaneously reducing overfitting \cite{jiang_deep_2020}. Moreover, one could take advantage of complementary deep learning and reinforcement learning based approaches such as importance sampling \cite{faury_improving_2019, sutton2018reinforcement}, or ``model-based'' methods that could utilize an electromagnetic surrogate model or inverse model to reduce the number of full electromagnetic Maxwell simulations needed for training \cite{pestourie_active_2020, hegde_deep_2020, kojima_deep_2021, tahersima_deep_2019}. Alternatively, whereas we optimized our grating for a single objective (single wavelength, $8^\circ$ scattering) Jiang and Fan showed that generative neural network based optimization can be extremely efficient for multi-objective design problems (e.g., multiple wavelengths and scattering angles in metasurfaces \cite{jiang_global_2019}). Along the same vein, in the next section we will show how a technique known as transfer learning can be used to re-purpose a neural network trained with PHORCED for an alternative objective, meanwhile boosting computational efficiency and electromagnetic performance dramatically.

\section{Transfer Learning with the PHORCED Method}
\label{sec:TL}

\begin{figure}[t!]
	\centering
	\includegraphics[width=6in]{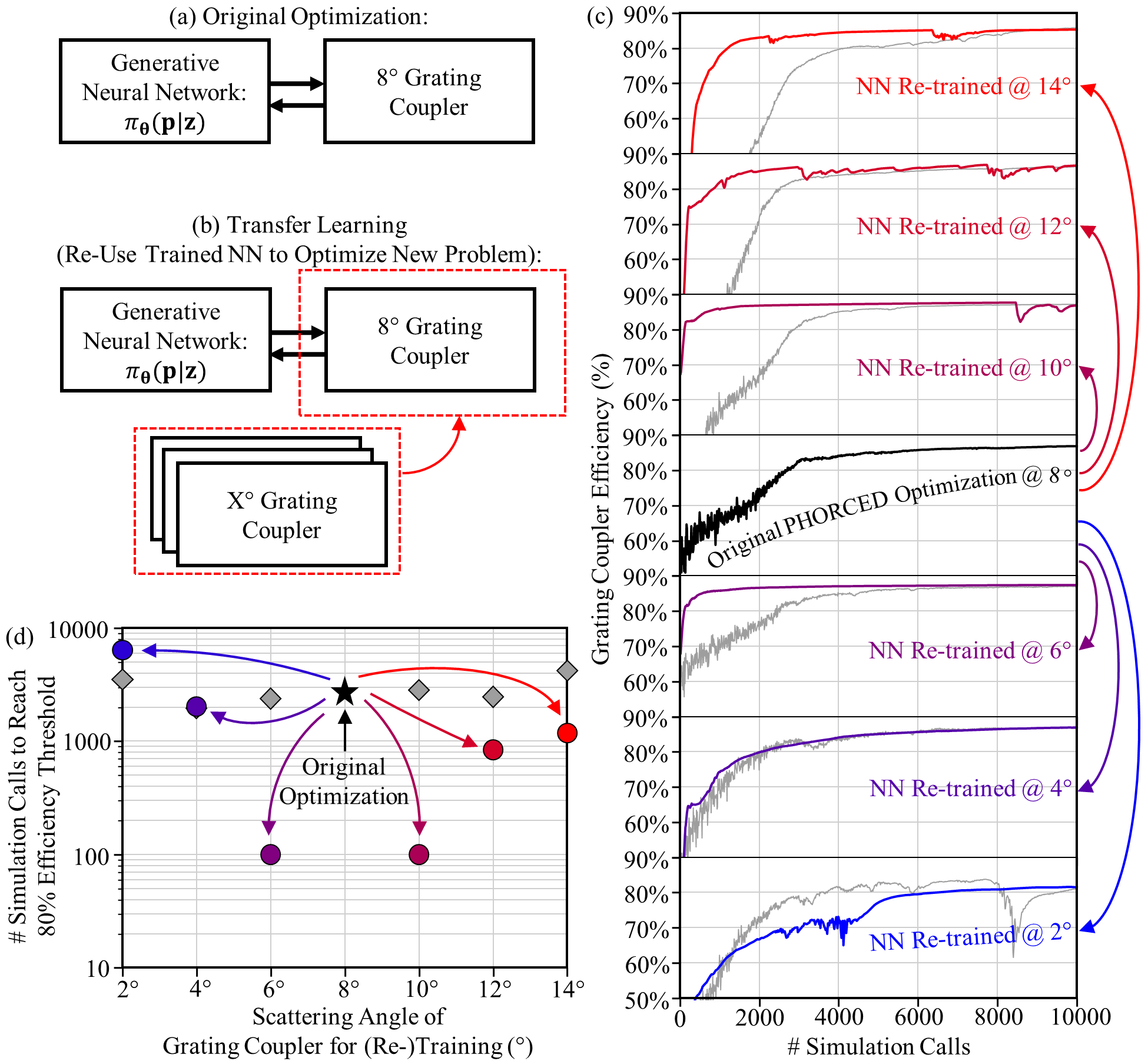}
	\caption{Transfer learning applied to grating coupler design yields accelerated convergence rate in optimization. The original PHORCED optimization from Fig.\,\ref{fig:fig2}(b) is qualitatively depicted as a block diagram in (a) for comparison with the transfer learning approach in (b). Here, we exchange the $8^\circ$ grating coupler merit function with an alternative grating coupler angle for re-training. Optimization progressions as a function of the number of simulation calls for each of the re-training sessions are shown in the blue/red curves of (c). Control optimizations where transfer learning was not applied are plotted in gray for comparison. In (d) we plot the number of simulation calls for each optimization from (c) to reach 80\% grating coupler scattering efficiency, with blue/red colored arrows and dots indicating applications of transfer learning and gray diamonds indicating the control cases.}
	\label{fig:fig3}
\end{figure}

Transfer learning is a concept in machine learning encompassing any method that reuses a model (e.g. a neural network) trained on one task for a different but related task. Qualitatively speaking and verified by real-world applications, we might expect the re-training of a neural network to occur faster than training a new model from scratch. Transfer learning has been extensively applied in classical machine learning tasks such as classification and regression, but has only recently been mentioned in the optics/photonics research domains \cite{qu_migrating_2019, jiang_deep_2020, hegde_deep_2020, ma_deep_2021}.  In this work we apply transfer learning to the inverse design of integrated photonics for the first time (to the authors' knowledge), revealing that a neural network trained using PHORCED for the design of $8^\circ$ grating couplers can be re-trained to design grating couplers with varied scattering angle and increased rate of convergence. 

\begin{figure}[t!]
    \centering
    \includegraphics[width=6in]{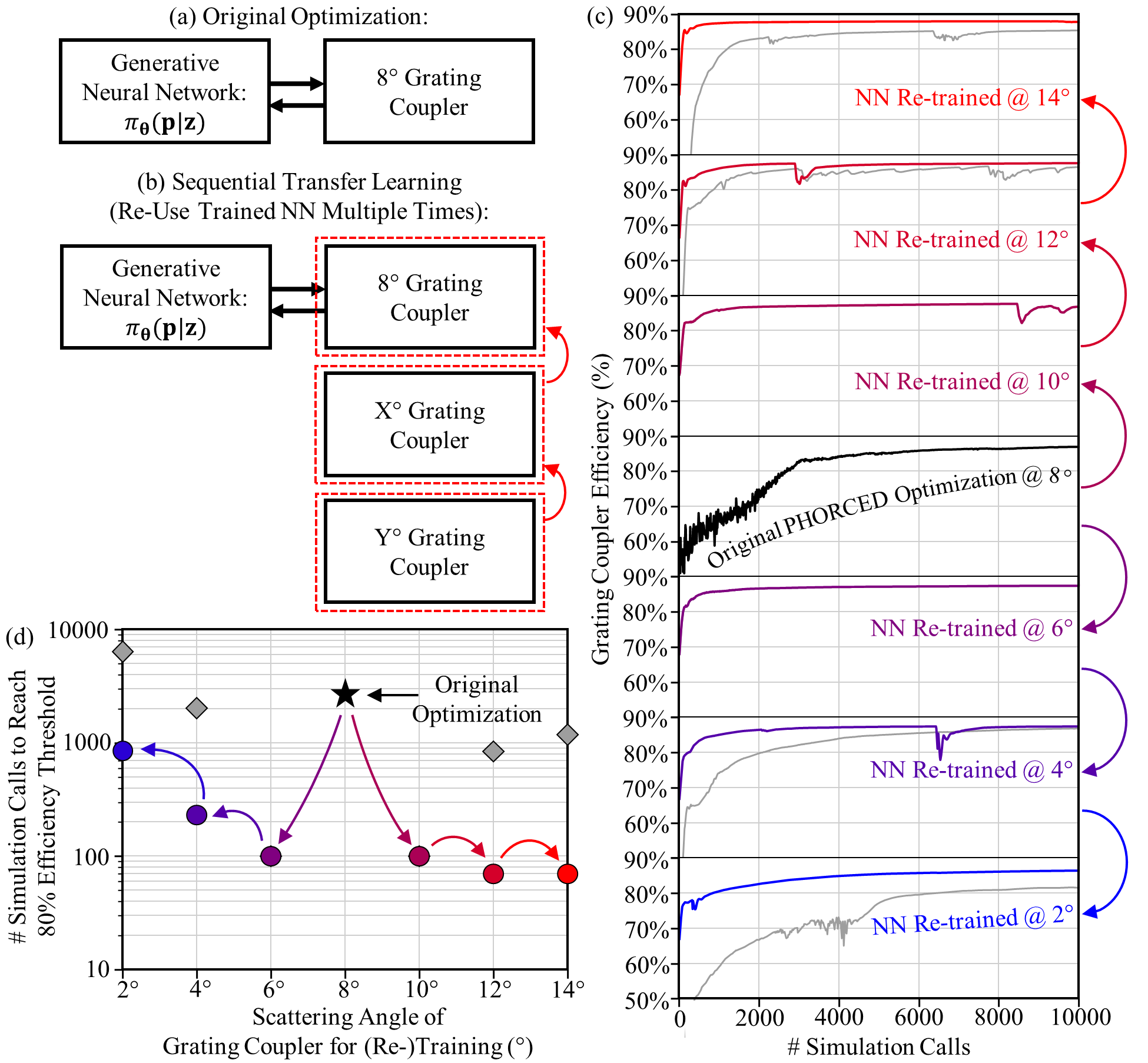}
    \caption{Transfer learning applied sequentially improves convergence for grating coupler scattering angles that are distant relative to the original optimization's scattering angle. The original PHORCED optimization from Fig.\,\ref{fig:fig2}(b) is qualitatively depicted in (a) for comparison with the ``sequential'' transfer learning approach in (b). Here, we sequentially re-train the neural network (originally trained to generate $8^\circ$ grating couplers) with progressively different scattering angles, with the intention of slowly changing the physics seen by the neural network. Grating coupler efficiency as a function of the number of simulation calls for each of the re-training sessions are shown in the blue/red curves of (c), where the arrows on the right-hand side show the sequence of each application of transfer learning. The results of the one-shot (non-sequential) transfer learning approach from Fig.\,\ref{fig:fig3}(c) are shown in gray for comparison. In (d) we plot the number of simulation calls needed for the grating coupler optimizations from (c) to reach an 80\% efficiency using the sequential transfer learning approach, similar to the corresponding plot in Fig.\,\ref{fig:fig3}(d). The blue/red arrows and dots indicate applications of sequential transfer learning, and the gray diamonds correspond to non-sequential transfer learning cases.}
    \label{fig:fig4}
\end{figure}

Transfer learning applied to grating coupler optimization is qualitatively illustrated in Fig.\,\ref{fig:fig3}(a)-(b). Fig.\,\ref{fig:fig3}(a) shows a shorthand version of the PHORCED optimization of Initial Seed 1 that was performed in Fig.\,\ref{fig:fig2}(b), where a neural network was specifically trained to design an $8^\circ$ grating coupler. In the case of transfer learning in Fig.\,\ref{fig:fig3}(b), we re-use the trained neural network from Fig.\,\ref{fig:fig3}(a) but now exchange the $8^\circ$ angle in the grating coupler efficiency merit function with an alternate scattering angle. In particular, we re-train the neural network on 6 alternative grating coupler angles: \{$2^\circ, 4^\circ, 6^\circ, 10^\circ, 12^\circ, 14^\circ$\}. Note that we maintained the exact same neural network architecture and optimization hyperparameters during these exchanges, including the optimization stopping criterion of 10,000 total simulation calls per training session; the only change in a given optimization was the grating coupler angle. As depicted in Fig.\,\ref{fig:fig3}(c), we show the optimization progressions of these transfer learning sessions (blue/red curves) in comparison to the original PHORCED optimization of the $8^\circ$ grating coupler from Fig.\,\ref{fig:fig2}(b) (reproduced in the black curve in the middle panel). Also shown are control optimizations for each grating coupler angle using the PHORCED method without transfer learning (in gray). We find that transfer learning to grating couplers with nearby scattering angles (e.g. $6^\circ$ and $10^\circ$) exhibit extremely accelerated rate of convergence relative to the original optimization and control cases. However, transfer learning is less effective or ineffective for more distant angles (e.g. $2^\circ$, $4^\circ$, and $14^\circ$). This observation is shown more clearly in Fig.\,\ref{fig:fig3}(d) where we plot the number of simulations required to reach 80\% efficiency in optimization versus the scattering angle for re-training \footnote{80\% grating coupler efficiency was chosen because it equates to roughly 1dB insertion loss -- an optimistic target for state-of-the-art silicon photonic devices.}. While the original optimization and control optimizations (black star and gray diamonds) required several thousand simulation calls before reaching this threshold, the $6^\circ$ and $10^\circ$ transfer learning optimizations required only about 100 simulations a piece -- a \textgreater$10\times$ reduction in simulation calls, making transfer learning comparable with local gradient-based optimization in terms of computation requirements. On the other hand, the distant grating coupler angle transfer learning optimizations ($2^\circ$, $4^\circ$, and $14^\circ$) required similar simulation call requirements to reach the same threshold as the original optimization. Evidently, there is a bias towards less effective transfer learning for small scattering angles. Grating couplers become plagued by parasitic back-reflection for small diffraction angles relative to normal \cite{michaels_inverse_2018,hooten_adjoint_2020}, and thus it is possible that the neural network has difficulty adapting to the new physics that were not previously encountered. We conclude that the transfer learning approach is most effective for devices with very similar physics to the device originally optimized by the neural network.

The results of Fig.\,\ref{fig:fig3} lead to a natural follow-up query: can we apply transfer learning multiple times progressively in order to maintain the convergence rate advantage for optimizations at more distant grating coupler angles? We explore this question of \textit{sequential} transfer learning in Fig.\,\ref{fig:fig4}. In Fig.\,\ref{fig:fig4}(a)-(b) we qualitatively compare sequential transfer learning to the original PHORCED optimization from Fig.\,\ref{fig:fig2}(b). As indicated, we replace the original $8^\circ$ grating coupler scattering angle in the electromagnetic merit function with an alternative scattering angle in the same manner discussed in Fig.\,\ref{fig:fig3}(b). Then, after that optimization has completed, we continue to iterate and exchange the grating coupler angle again. By sequentially applying transfer learning, we hope to slowly introduce new physics to the neural network such that we can maintain faster convergence at more physically distant problems from the initial optimization. We conduct two sequential transfer learning sessions where we evolve the grating coupler scattering angle in the following steps: $\{8^\circ \rightarrow 10^\circ \rightarrow 12^\circ\rightarrow 14^\circ\}$ and $\{8^\circ \rightarrow 6^\circ \rightarrow 4^\circ \rightarrow 2^\circ\}$. The results of these progressions are shown in Fig.\,\ref{fig:fig4}(c) where we plot the grating coupler efficiency as a function of the number of simulation calls in each (re-)training session. The $6^\circ$, $8^\circ$, and $10^\circ$ cases are the same as those shown previously in Fig.\,\ref{fig:fig3}(c); the new results may be seen in the $2^\circ$, $4^\circ$, $12^\circ$, and $14^\circ$ cases, where blue/red lines indicate the new optimization data and gray lines indicate the non-sequential transfer learning cases from Fig.\,\ref{fig:fig3}(c) for comparison. We observe that sequential transfer learning improves the optimization convergence rate for the distant grating coupler scattering angles, in accordance with our initial prediction. This observation is made more explicit in Fig.\,\ref{fig:fig4}(d) where we plot the simulation call requirement to reach an 80\% grating coupler efficiency threshold for each of the transfer learning optimizations. Blue/red arrows and points indicate sequential transfer learning progressions, while gray diamonds indicate the one-shot transfer learning cases reproduced from Fig.\,\ref{fig:fig3}(d). Evidently, sequential transfer learning improves simulation call requirements by approximately an order of magnitude relative to the single-step cases. While we find that there is still a noticeable bias towards longer (but less severely long) training at smaller scattering angles, sequential transfer learning had the added benefit of providing the most efficient devices overall at every scattering angle in the progression. Note that the plotted simulation call requirements do not include the simulation calls from the previous iteration in the sequential transfer learning progression (each application of transfer learning used 10,000 simulation calls, where the final neural network weightset after those 10,000 simulation calls was used as the initial weightset for the next iteration of transfer learning). Furthermore, note that each optimization in these sequential transfer learning cases used the same neural network architecture and hyperparameters as the original PHORCED optimization ($8^\circ$ case), except for the $\{12^\circ \rightarrow 14^\circ\}$ and $\{4^\circ \rightarrow 2^\circ\}$ cases which required a slightly smaller learning rate in optimization for better performance. The smaller learning rate negligibly affected the 80\% efficiency simulation call requirement shown in Fig.\,\ref{fig:fig4}(d).

\section{Conclusion}
In this work we introduced PHORCED, a photonic optimization package leveraging the policy gradient method from reinforcement learning. This method interfaces a probablistic neural network with an electromagnetic solver for enhanced inverse design capabilities. PHORCED does not require an evaluation of adjoint method gradient of the electromagnetic merit function with respect to the design parameters, therefore eliminating the need to perform adjoint simulations over the course of an optimization. We anticipate that this fact can be particularly advantageous for multi-frequency electromagnetic merit functions, where multiple adjoint simulations would normally be required in a simple frequency-domain implementation of the adjoint method (e.g., see Ref.\,\cite{hooten_adjoint_2020}).  

We applied both PHORCED and the GLOnet method to the proof-of-concept optimization of grating couplers. We found that both algorithms could outperform conventional gradient-based BFGS optimization, resulting in state-of-the-art simulated insertion loss for single-etch c-Si grating couplers and resilience against poor choices of initial condition. In future work we intend to implement fabrication constraints, alternative choices of geometrical parameterization, and other criteria to guarantee feasibility and robustness of experimental devices.

As an additional contribution we introduced the concept of transfer learning to integrated photonic optimization, revealing that a trained neural network using PHORCED could be re-trained on alternative problems with accelerated convergence. In particular, we showed that transfer learning could be applied to the design of grating couplers with varied scattering angle. Transfer learning was extremely effective for grating coupler scattering angles within $\approx\pm4^\circ$ to the original optimization angle, improving the convergence rate by \textgreater $10\times$ in some cases. However, this range could be effectively extended to $\approx \pm 6^\circ$ or more using a sequential transfer learning approach, where transfer learning was applied multiple times progressively to slowly change the angle seen by the neural network. Because neural network based design methods such as PHORCED are generally data-hungry, we believe that transfer learning could greatly reduce the electromagnetic simulation and compute time that would otherwise be required by these techniques in the design of complex electromagnetic structures. For example, transfer learning could be used in multiple hierarchical stages to evolve an optimization from a two-dimensional structure to a three-dimensional structure, or from a surrogate model (e.g., the grating coupler model in Ref.\,\cite{gostimirovic_open-source_2019}) to real physics.

Looking forward, we would like to emphasize that PHORCED takes advantage of fundamental concepts in reinforcement learning, but there is a plethora of burgeoning contemporary research in this field, such as advanced policy gradient, off-policy, and model-based approaches. We anticipate that further cross-pollination of the inverse electromagnetic design and reinforcement learning communities could open the floodgates for new research in electromagnetic optimization.

\printbibliography


%
%



\end{document}


\maketitle

\renewcommand{\theequation}{S.\arabic{equation}}
\renewcommand\thefigure{S\arabic{figure}}

\section{Theory of GLOnet and PHORCED Optimization Algorithms}
Below we will explain and derive the formulae presented in Fig. 1 of the main manuscript, particularly the gradient/backpropagation terms. The primary intention of this section is to justify the claim that PHORCED does not require an adjoint simulation. This section will assume the reader has a basic understanding of gradient-based optimization and the adjoint method. The reader is recommended to consult Refs.\,\cite{michaels_inverse_2018, su_fully-automated_2017, hooten_adjoint_2020, sun_adjoint-method-inspired_2020, dezfouli_design_2021} for more detailed explanation of these topics.

In the following, let $\mathbf{p}=[p_1, p_2,...,p_{n-1},p_n]^T$ be a vector of $n$ geometrical degrees-of-freedom (parameters) that we wish to vary in our device. Furthermore, the time-harmonic Maxwell's Equations may be written as a matrix equation:
\begin{align}
    \text{Maxwell's Equations: }\mathbf{A}\mathbf{x}=\mathbf{b}\label{eq:ax_b}
\end{align}
with:
\begin{align}
        \mathbf{A}=\begin{bmatrix}
            j\omega\varepsilon(\mathbf{p}) &  \nabla \times \\
            \nabla \times & -j\omega \mu(\mathbf{p})
        \end{bmatrix}, \quad
        \mathbf{x}=\begin{bmatrix}
            \mathbf{E} \\
            \mathbf{H}
        \end{bmatrix}, \quad
        \mathbf{b}=\begin{bmatrix}
            \mathbf{J}_e \\
            \mathbf{J}_m
        \end{bmatrix} 
\end{align}
where $j$ is the imaginary unit, $\omega$ is the frequency, $\mathbf{A}$ is the Maxwell Operator, $\mathbf{x}$ is a vector of electric and magnetic fields $\mathbf{E}$ and $\mathbf{H}$, and $\mathbf{b}$ is a vector of electric and magnetic current sources $\mathbf{J}_e$ and $\mathbf{J}_m$. Notice that $\mathbf{A}$ includes the permittivity and permeability $\varepsilon$ and $\mu$ which are explicitly parameterized by the geometrical degrees-of-freedom $\mathbf{p}$. We will assume that the electric and magnetic field vectors each have dimension $k$, which might be the total number of Yee cells in an FDTD simulation. Note also that the electric and magnetic fields are, in general, complex-valued.

A solution to Maxwell's Equations (e.g. a simulation), can be regarded as solving Eq.\,\eqref{eq:ax_b} for $\mathbf{x}$ given $\mathbf{A}$ and $\mathbf{b}$. A shorthand way to write this is:
\begin{align}
    \mathbf{x} = \mathbf{A}^{-1}\mathbf{b}\label{eq:ax_b_solve}
\end{align}
Though, it should be emphasized that taking an explicit inverse of $\mathbf{A}$ is rarely done in practice (and not to mention, not well-defined on a continuous domain). Since $\mathbf{A}$ is a function of $\mathbf{p}$, we will rewrite Eq.\,\eqref{eq:ax_b_solve} in a more convenient form. Without loss of generality, let $g:\mathcal{R}^n\rightarrow \mathcal{C}^{2k}$ be a function that maps geometrical degrees-of-freedom to electric and magnetic field quantities. Then, we can write:
\begin{align}
    \boxed{\text{Forward Simulation: } \mathbf{x} = g(\mathbf{p})}\label{eq:x_gp}
\end{align}
which represents the operation given above in Eq.\,\eqref{eq:ax_b_solve}, but with explicit dependence on $\mathbf{p}$. Please note that $g$ is specific to a particular electromagnetic solution of Maxwell's equations with implicitly-defined boundary conditions and geometrical mapping of the permittivity and permeability; one should always refer back to Eq.\,\eqref{eq:ax_b} for the most general solution to Maxwell's Equations.

The last requirement for a well-defined inverse design optimization problem is a merit function, which defines the performance or success of a given electromagnetic device. Let $f:\mathcal{C}^{2k}\rightarrow\mathcal{R}$ be a function that maps electric and magnetic field quantities to a performance metric (figure-of-merit). Then, given $\mathbf{x}=g(\mathbf{p})$ for given parameters $\mathbf{p}$ we may write:
\begin{align}
    \text{Figure-of-Merit}&=f(\mathbf{x}) \\
    \text{Figure-of-Merit}&=(f\circ g)(\mathbf{p})\label{eq:fom_f_g_p}
\end{align}
where in Eq.\,\eqref{eq:fom_f_g_p} we made use of Eq.\,\eqref{eq:x_gp} to write the total figure-of-merit as a function composition of the electromagnetic merit function and the solution to Maxwell's Equations given well-defined $\mathbf{p}$. The intention of an inverse design optimization is to finding the best set of geometrical degrees-of-freedom that improve the figure-of-merit. Assuming that we intend to maximize the figure-of-merit, we may write a generalized electromagnetic inverse design problem in optimization notation as the following:
\begin{align}
    \boxed{\text{Inverse Design Optimization: } \mathbf{p}^* = \argmax_{\mathbf{p}}{(f\circ g)(\mathbf{p})}}\label{eq:p_inv}
\end{align}
where $\mathbf{p}^*$ is the global optimum of $(f\circ g)$. Using optimization methods such as gradient-descent, GLOnet, and PHORCED, we hope to attain $\mathbf{p}^*$ in a computationally efficient way.

Given the convenient notation in Eqs.\,\eqref{eq:x_gp}-\eqref{eq:p_inv}
we are now prepared to explain the three optimization algorithms given in Fig.\,1 of the main text. We begin with simple gradient-based optimization.

\subsection{Gradient-Based Optimization}
In gradient-based optimization we intend to use the gradient of a function to inform an optimization update step. In terms of our inverse design optimization problem Eq.\,\eqref{eq:p_inv}, we take the gradient of the figure-of-merit with respect to the geometrical variable $\mathbf{p}$ defined at some given $\mathbf{p}'$. For convenience we write this gradient in a slightly shorthand way:
\begin{align}
    \text{Figure-of-Merit Gradient: }\frac{\partial(f\circ g)}{\partial \mathbf{p}'} \equiv \nabla_{\mathbf{p}} \left[(f\circ g)(\mathbf{p})\right]_{\mathbf{p}=\mathbf{p}'}\label{eq:gradient}
\end{align}
In electromagnetic problems, the gradient is most efficiently calculated using the adjoint method. An exact derivation of the adjoint method may be found in Refs.\,\cite{michaels_inverse_2018, su_fully-automated_2017, hooten_adjoint_2020, sun_adjoint-method-inspired_2020, dezfouli_design_2021}. For our purposes, we require only the qualitative understanding that for any given $\mathbf{p}'$, we may solve Eq.\,\eqref{eq:gradient} using just 2 simulations (the forward and adjoint simulations respectively), regardless of the dimension of the parameter vector, $n=\text{dim}(\mathbf{p}')$. Using this gradient, we can then update the parameter vector that will be used in the subsequent iteration of our gradient-based optimization algorithm. In this work we used the Broyden–Fletcher–Goldfarb–Shanno (BFGS) algorithm for our baseline gradient-based optimization comparisons. This algorithm was implemented out-of-the-box from the scipy.optimize python module, where forward/adjoint simulations and gradient calculations were performed in the open-source electromagnetic optimization package EMopt \cite{michaels_emopt_2019}.

\subsection{GLOnet}
In this section we will derive the backpropagation term from Fig.\,1 of the main manuscript. Note that this derivation differs from that originally given by Jiang and Fan \cite{jiang_global_2019,jiang_simulator-based_2020}, we refer the reader there for more detail and a for a different perspective from that discussed here.  Let $\mathbf{z}\sim \mathcal{D}$ be a random vector drawn from a ($d$-dimensional) distribution $\mathcal{D}$. Importantly, this noise vector serves as an input for a neural network that deterministically generates geometrical parameter vectors for simulation, based on programmable weights denoted by $\theta$. Specifically, let $h_{\boldsymbol{\theta}}: \mathcal{R}^d\rightarrow\mathcal{R}^n$ represent this function such that for given $\boldsymbol{\theta}$ and sampled $\mathbf{z}\sim\mathcal{D}$ we have:
\begin{align}
    \mathbf{p} = h_{\boldsymbol{\theta}} (\mathbf{z})
\end{align}
By virtue of $\mathbf{z}$ being a random variable, $\mathbf{p}$ is also random but drawn from an unknown distribution parameterized by $\boldsymbol{\theta}$. Consequently, the objective function for inverse design optimization via GLOnet becomes:
\begin{align}
    (f\circ g)(\mathbf{p}) \rightarrow \EX_{\mathbf{p}=h_{\boldsymbol{\theta}} (\mathbf{z})} [(f\circ g)(\mathbf{p})] = \EX_{\mathbf{z}\sim \mathcal{D}} [(f\circ g \circ h_{\boldsymbol{\theta}})(\mathbf{z})]
\end{align}
where $\EX[\cdot]$ is the expected value operator. In the right-hand side we simplified the expression by replacing $\mathbf{p}=h_{\boldsymbol{\theta}} (\mathbf{z})$. Intuitively our intention is to optimize the average value of the electromagnetic merit function generated by the neural network. Note that because of this exchange, the variable(s) that may be explicitly controlled by the user or optimization algorithm are the programmable weights, $\boldsymbol{\theta}$. Consequently, the optimization problem for GLOnet may be written:
\begin{align}
    \boxed{\text{GLOnet Optimization: } \boldsymbol{\theta}^* = \argmax_{\boldsymbol{\theta}} \EX_{\mathbf{z}} [(f\circ g \circ h_{\boldsymbol{\theta}})(\mathbf{z})]}\label{eq:glonet}
\end{align}
where we wrote $\mathbf{z}$ as the variable of the expected value with the understanding that it is sampled from user-defined distribution $\mathcal{D}$. To optimize $\boldsymbol{\theta}$ for this objective, we will leverage the powerful neural network optimization package called PyTorch \cite{paszke2017automatic}. This tool allows one to invoke state-of-the-art gradient-based algorithms for the optimization of neural network weights. This will require the partial derivatives of the GLOnet objective with respect to the neural networks weights. 

Let the probability density function of $\mathbf{z}$ be denoted $\text{pdf}(\mathbf{z})$, then we may write:
\begin{align}
    \EX_{\mathbf{z}} [(f\circ g \circ h_{\boldsymbol{\theta}})(\mathbf{z})] = \int (f\circ g \circ h_{\boldsymbol{\theta}})(\mathbf{z}) \text{pdf}(\mathbf{z})d\mathbf{z}
\end{align}
Then for scalar weight $\theta_j\in\boldsymbol{\theta}$ in our neural network, the partial derivative of the objective is given by:
\begin{align}
    \frac{\partial}{\partial \theta_j}\EX_{\mathbf{z}} [(f\circ g \circ h_{\boldsymbol{\theta}})(\mathbf{z})] = \int \frac{\partial (f\circ g \circ h_{\boldsymbol{\theta}})(\mathbf{z})}{\partial \theta_j} \text{pdf}(\mathbf{z})d\mathbf{z}
\end{align}
where we invoked the Leibniz integral rule for the exchange of the integral and partial derivative operators. Let $\mathbf{p}=h_{\boldsymbol{\theta}}(\mathbf{z})$ be the output of the neural network for input $\mathbf{z}$. Then we may apply the chain rule to the derivative term:
\begin{align}
    \frac{\partial}{\partial \theta_j}\EX_{\mathbf{z}} [(f\circ g \circ h_{\boldsymbol{\theta}})(\mathbf{z})] &= \int \frac{\partial (f\circ g)}{\partial \mathbf{p}}\cdot\frac{\partial \mathbf{p}}{\partial \theta_j} \text{pdf}(\mathbf{z})d\mathbf{z} \\
    &= \EX_\mathbf{z}\left[\frac{\partial (f\circ g)}{\partial \mathbf{p}}\cdot\frac{\partial \mathbf{p}}{\partial \theta_j} \bigg|_{\mathbf{p}=h_{\boldsymbol{\theta}}(\mathbf{z})}\right]
\end{align}
where in the second step we re-wrote the integral as an expected value. Note that the ``$\cdot$'' operator is the vector dot product. Abusing notation slightly, we may write the full gradient as:
\begin{align}
   \frac{\partial}{\partial \boldsymbol{\theta}}\EX_{\mathbf{z}} [(f\circ g \circ h_{\boldsymbol{\theta}})(\mathbf{z})]&= \EX_\mathbf{z}\left[\frac{\partial (f\circ g)}{\partial \mathbf{p}}\cdot\frac{\partial \mathbf{p}}{\partial \boldsymbol{\theta}} \bigg|_{\mathbf{p}=h_{\boldsymbol{\theta}}(\mathbf{z})}\right]\label{eq:glonet_deriv}
\end{align}
which concludes the proof. Observe that the first expression within the expected value is simply the adjoint method gradient described previously in Eq.\,\eqref{eq:gradient}, and therefore requires two simulations per output vector $\mathbf{p}$ of the neural network. The additional term $\partial \mathbf{p}/\partial \boldsymbol{\theta}$ may be evaluated using automatic differentiation.  

Note that in the original paper by Jiang and Fan \cite{jiang_global_2019,jiang_simulator-based_2020}, the authors emphasized taking the exponential value of the electromagnetic merit function, $f\rightarrow \exp\{f/\sigma\}$, where $\sigma$ is a hyperparameter. We can include this additional contribution explicitly in Eq.\,\eqref{eq:glonet} and Eq.\,\eqref{eq:glonet_deriv} through an application of the chain rule:
\begin{align}
    \boxed{
    \text{GLOnet Objective: }\EX_{\mathbf{z}}\left[\exp \left\{\frac{(f\circ g \circ h_{\boldsymbol{\theta}})(\mathbf{z})}{\sigma}\right\}\right]}\label{eq:glonet_obj_summ}
\end{align}

\begin{align}
    \boxed{
    \text{Backpropagation Gradient: }\EX_\mathbf{z}\left[\frac{1}{\sigma}\exp\left\{\frac{(f\circ g)(\mathbf{p})}{\sigma}\right\}\frac{\partial (f\circ g)}{\partial \mathbf{p}}\cdot\frac{\partial \mathbf{p}}{\partial {\boldsymbol{\theta}}} \bigg|_{\mathbf{p}=h_{\boldsymbol{\theta}}(\mathbf{z})}\right]}
    \label{eq:glonet_deriv_summ}
\end{align}

In the main text the exponential term was excluded for clarity, but for the GLOnet results in Fig.\,2 this is the form of the objective that was used, where we chose hyperparameter $\sigma=0.6$.

\subsection{PHORCED}
As mentioned the main text, PHORCED is qualitatively similar to GLOnet in the sense that we wish to use a generative neural network to suggest geometrical degrees-of-freedom. However, by contrast to GLOnet where $\mathbf{p}$ is provided deterministically from the neural network, in PHORCED we treat $\mathbf{p}$ as a random vector sampled from a conditional probability distribution, $\pi_{\boldsymbol{\theta}}$:
\begin{align}
    \mathbf{p}\sim\pi_{\boldsymbol{\theta}}(\mathbf{p}|\mathbf{z})
\end{align}
where $\mathbf{z}\sim \mathcal{D}$ is once again an input vector, which in this case conditions the distribution. Note that, for given $\mathbf{z}$, the distribution defined by $\pi_{\boldsymbol{\theta}}$ is not static and is in fact programmable by virtue of the neural network weights ${\boldsymbol{\theta}}$ that determine its statistical parameters. In the main text of this paper we chose $\pi_{\boldsymbol{\theta}}(\mathbf{p}|\mathbf{z})$ to be a multivariate Gaussian, with mean and covariance matrix defined by a generative neural network with weights ${\boldsymbol{\theta}}$. 

Because $\mathbf{p}$ is now explicitly a random vector (not random just by virtue of $\mathbf{z}$ being random) the objective of PHORCED is modified relative to GLOnet (from Eq.\,\eqref{eq:glonet}):
\begin{align}
    \boxed{\text{PHORCED Optimization: } {\boldsymbol{\theta}}^* = \argmax_{\boldsymbol{\theta}} \EX_{(\mathbf{p},\mathbf{z})}[(f\circ g)(\mathbf{p})]}
\end{align}
where the subtle difference is we now wish to improve the expected value of the electromagnetic merit function under the joint probability of sampling random vectors $\mathbf{p}$ and $\mathbf{z}$ \footnote{In the general reinforcement learning literature, the merit function within the expected value may be a function of both $\mathbf{p}$ and $\mathbf{z}$, but this situation was excluded because it is non-applicable here.}. As a consequence of this, we note that we may write the joint probability density function of $(\mathbf{p},\mathbf{z})$ as:
\begin{align}
    \text{pdf}(\mathbf{p},\mathbf{z})=\pi_{\boldsymbol{\theta}} (\mathbf{p}|\mathbf{z})\text{pdf}(\mathbf{z})
\end{align}
Hence,
\begin{align}
    \EX_{(\mathbf{p},\mathbf{z})}[(f\circ g)(\mathbf{p})]&=\iint (f\circ g)(\mathbf{p}) \text{pdf}(\mathbf{p},\mathbf{z})d\mathbf{p}d\mathbf{z} \\
    &=\iint (f\circ g)(\mathbf{p}) \pi_{\boldsymbol{\theta}}(\mathbf{p}|\mathbf{z})\text{pdf}(\mathbf{z})d\mathbf{p}d\mathbf{z}
\end{align}
Then for scalar weight $\theta_j\in \boldsymbol{\theta}$ in our neural network, the partial derivatives of this expected value are given by:
\begin{align}
    \frac{\partial}{\partial \theta_j}\EX_{(\mathbf{p},\mathbf{z})}[(f\circ g)(\mathbf{p})]&=\iint (f\circ g)(\mathbf{p}) \frac{\partial \pi_{\boldsymbol{\theta}}(\mathbf{p}|\mathbf{z})}{\partial \theta_j}\text{pdf}(\mathbf{z})d\mathbf{p}d\mathbf{z}\label{eq:phorced_pre_log}
\end{align}
where we applied the Leibniz integral rule followed by the product rule. Observe the very important fact that neither $\mathbf{p}$ nor $(f\circ g)$ are explicitly related to $\theta_j$, and therefore a term of the form $\frac{\partial(f\circ g)(\mathbf{p})}{\partial \theta_j}$ is zero in the product rule derivative. Indeed, only the policy distribution, $\pi_{\boldsymbol{\theta}}$, is modeled by the neural network, and therefore is the only quantity subject to the derivative. Moreover, we note that by the ''log trick'' we may write:
\begin{align}
    \frac{\partial \pi_{\boldsymbol{\theta}}(\mathbf{p}|\mathbf{\mathbf{z})}}{\partial \theta_j} = \pi_{\boldsymbol{\theta}}(\mathbf{p}|\mathbf{z})\frac{\partial}{\partial \theta_j} \log \pi_{\boldsymbol{\theta}}(\mathbf{p}|\mathbf{z})\label{eq:log_trick}
\end{align}
to which Eq.\,\eqref{eq:phorced_pre_log} becomes:
\begin{align}
    \frac{\partial}{\partial \theta_j}\EX_{(\mathbf{p},\mathbf{z})}[(f\circ g)(\mathbf{p})]&=\iint (f\circ g)(\mathbf{p}) \frac{\partial \log \pi_{\boldsymbol{\theta}}(\mathbf{p}|\mathbf{z})}{\partial \theta_j}\pi_{\boldsymbol{\theta}}(\mathbf{p}|\mathbf{z})\text{pdf}(\mathbf{z})d\mathbf{p}d\mathbf{z} \\
    &=\EX_{(\mathbf{p},\mathbf{z})}\left[(f\circ g)(\mathbf{p})  \frac{\partial \log \pi_{\boldsymbol{\theta}}(\mathbf{p}|\mathbf{z})}{\partial \theta_j}\right]\label{eq:phorced_post_log_trick}
\end{align}
where in the second step we used Eq.\,\eqref{eq:log_trick} and the definition of the expected value. Eq.\,\eqref{eq:phorced_post_log_trick} was reported in the main text as the quantity used for backpropagation. In our implementation we also included a ``baseline subtraction'' term, where we subtract the sample average of the electromagnetic merit function, $\text{average}_{\mathbf{p}}[(f\circ g)(\mathbf{p})]$, from the electromagnetic merit function each iteration of the optimization routine. This heuristic is well-known in the reinforcement learning to reduce model variance without affecting bias in expectation (Ref.\,\cite{sutton2018reinforcement}).

To summarize, the objective and backprogation terms for PHORCED are given by,
\begin{align}
    \boxed{\text{PHORCED Objective: } \EX_{(\mathbf{p},\mathbf{z})}\left[(f\circ g)(\mathbf{p})\right]}
\end{align}

\begin{align}
    \boxed{\text{Backpropagation Gradient: } \EX_{(\mathbf{p},\mathbf{z})}\left[\big((f\circ g)(\mathbf{p})-b \big)  \frac{\partial \log \pi_{\boldsymbol{\theta}}(\mathbf{p}|\mathbf{z})}{\partial \boldsymbol{\theta}}\right]}
\end{align}
\begin{align}
    \boxed{b\approx \EX_{(\mathbf{p},\mathbf{z})}\left[(f\circ g)(\mathbf{p})\right]}
\end{align}
Observe that PHORCED requires no evaluation of the gradient of the electromagnetic figure of merit: $\frac{\partial (f\circ g)}{\partial \mathbf{p}}$. Indeed, the foremost term within the gradient expression requires only ``forward simulation'' evaluations. Meanwhile, the latter term $\partial \log \pi_{\boldsymbol{\theta}}/\partial \boldsymbol{\theta}$ may be computed using pytorch's automatic differentiation feature. These features of PHORCED should be contrasted with the similar equations from GLOnet (Eqs.\,\eqref{eq:glonet_obj_summ}-\eqref{eq:glonet_deriv_summ}), where we note that the optimization objective is identical but the gradient required for backpropagation changes drastically by our representation of the neural network model.

\section{Neural Network Model Specifications}
\begin{figure}[h!]
    \centering
    \includegraphics[width=13cm]{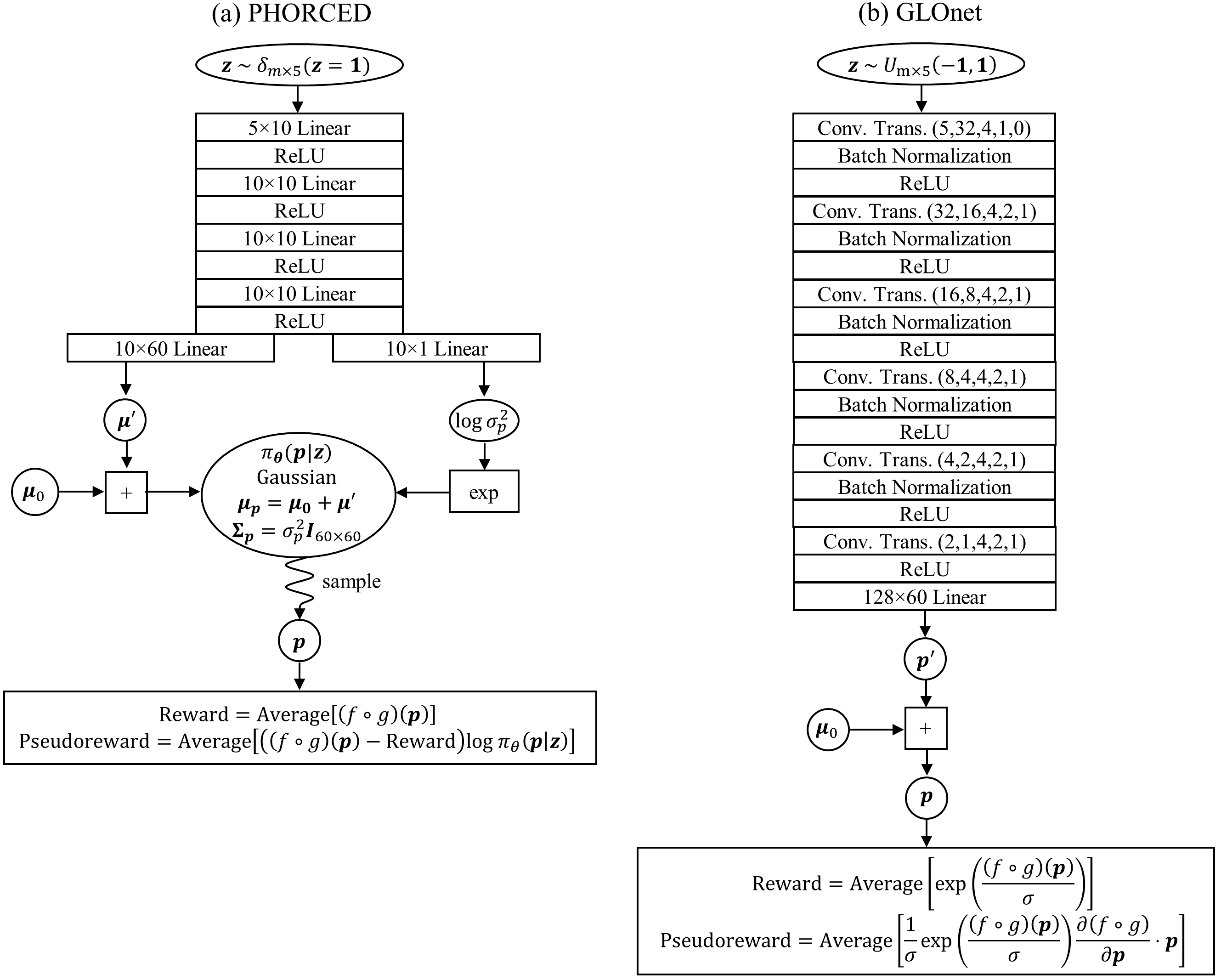}
    \caption{Neural network models used for optimization with (a) PHORCED and (b) GLOnet. Note that ``Reward'' is the objective of interest, whereas the ``Pseudoreward'' is the objective value actually seen by the neural network. This allows us to conveniently invoke automatic differentiation despite performing electromagnetic simulations and gradient calculations outside of PyTorch's autograd framework. Other definitions and hyperparameters may be found in Fig.\,\ref{fig:defs}.}
    \label{fig:nets}
\end{figure}

\begin{figure}[h!]
    \centering
    \includegraphics[width=8cm]{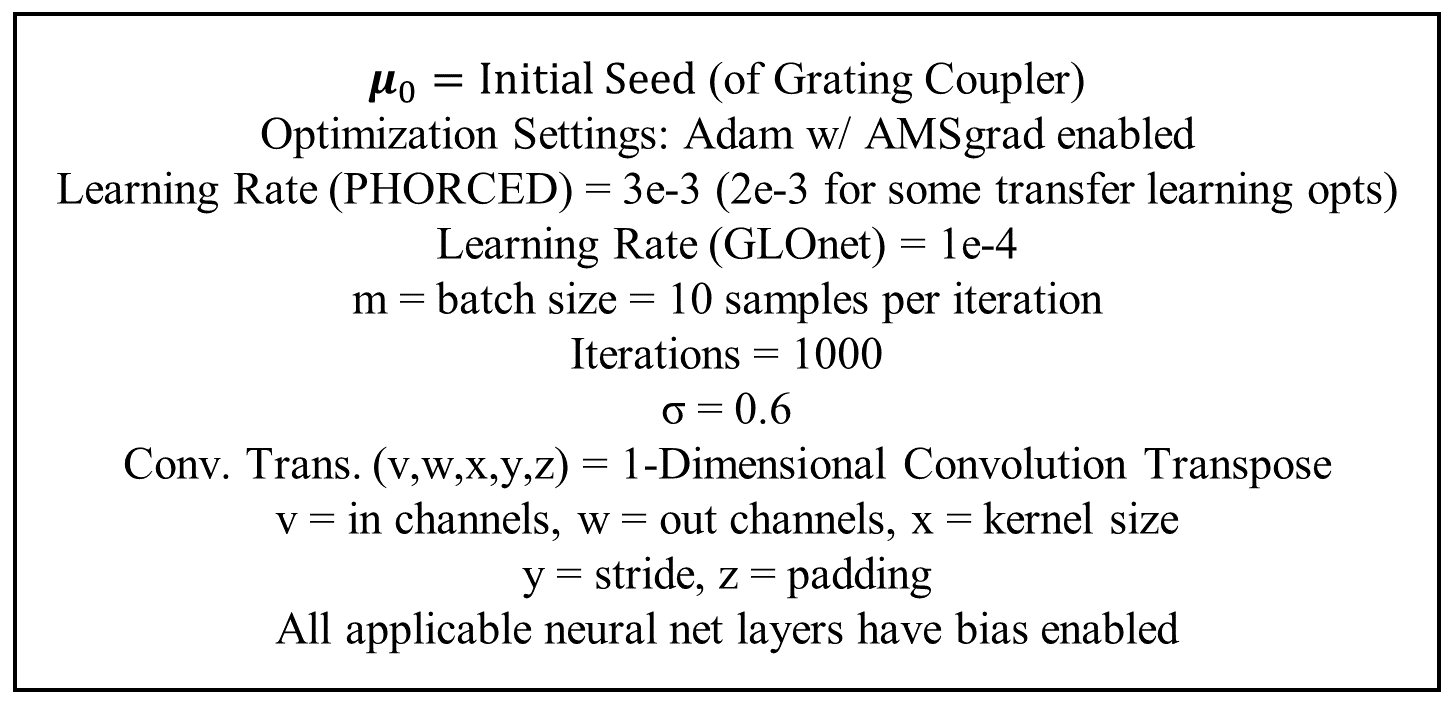}
    \caption{Definitions and hyperparameter specifications.}
    \label{fig:defs}
\end{figure}

In Fig.\,\ref{fig:nets} we illustrate the architectures and hyperparameters for the implementation of the generative neural network models used to obtain the results in Figs. 2-4 from the main text. Both PHORCED and GLOnet are used to output design vectors of dimension 60, representing the width and spacing between 30 grating etches. Note that there are better performing choices of grating coupler etch parameterization \cite{michaels_inverse_2018, hooten_adjoint_2020}, but this parameterization was chosen to illustrate how the algorithm would behave in high-dimensional situations where less physical intuition is available. Both models were implemented in PyTorch and electromagnetic simulations were performed in EMopt \cite{michaels_emopt_2019}. Hyperparameter definitions are provided in Fig.\,\ref{fig:defs}. Note that we were unable to perform exhaustive hyperparameter testing in this work due to the prohibitively slow speed of electromagnetic simulations of this size. Training with PHORCED and GLOnet required \textasciitilde 12 and \textasciitilde 36 hours on a high-performance server.

\newpage

\printbibliography

%
%

